\documentclass[le]{aa}  
\usepackage{graphicx}
\usepackage{txfonts}
\usepackage{float}
\begin{document}

   \title{A search for stellar tidal streams around Milky Way analogues from the SAGA sample}

%   \subtitle{I. Overviewing the $\kappa$-mechanism}

   \author{Juan Mir\'{o}-Carretero \inst{1}, David Mart\' \i nez-Delgado\inst{3}, S\'{i}lvia Farr\`{a}s-Aloy \inst{2,9}, Maria A. G\'{o}mez-Flechoso \inst{1,10}, Andrew Cooper \inst{4,5,11}, Santi Roca-F\`{a}brega\inst{1,6}, Konrad Kuijken \inst{9}, Mohammad Akhlaghi \inst{7}, Giussepe Donatiello \inst{8}
   %Dustin Lang \inst{9}
   }

   \institute{Departamento de F{\'\i}sica de la Tierra y Astrof{\'\i}sica, Universidad Complutense de Madrid, Plaza de las Ciencias, E-28040 Madrid, Spain
        \and
            Universidad Internacional de Valencia (VIU), C. del Pintor Sorolla 21, 46002 Valencia, Spain  
         \and
             Instituto de Astrof\'isica de Andaluc\'ia, CSIC, Glorieta de la Astronom\'\i a, E-18080, Granada, Spain 
        \and
        Institute of Astronomy and Department of Physics, National Tsing Hua University, Kuang Fu Rd. Sec. 2, Hsinchu 30013, Taiwan
        \and
Center for Informatics and Computation in Astronomy, National Tsing Hua University, Kuang Fu Rd. Sec. 2, Hsinchu 30013, Taiwan
\and
Instituto de Astronomía, Universidad Nacional Autónoma de México, Apartado Postal 106, C. P. 22800, Ensenada, B. C., Mexico 
         \and
             Centro de Estudios de Física del Cosmos de Aragón (CEFCA), Plaza San Juan 1, 44001 Teruel, Spain
         \and
         UAI - Unione Astrofili Italiani /P.I. Sezione Nazionale di Ricerca Profondo Cielo, 72024 Oria, Italy
        %\and
        %     Perimeter Institute for Theoretical %Physics, 31 Caroline St N, Waterloo, %Canada
        \and
             Leiden Observatory, Leiden University, Niels Bohrweg 2, NL-2333 CA Leiden, the Netherlands
        \and
             Instituto de Física de Partículas y del Cosmos (IPARCOS), Fac. CC. Físicas, Universidad Complutense de Madrid, Plaza de las Ciencias, 1, E-28040 Madrid, Spain
        \and 
             Physics Division, National Center for Theoretical Sciences, Taipei 10617, Taiwan
             }
\titlerunning{A search for stellar tidal streams around Milky Way analogues from the SAGA sample}
\authorrunning{Mir\'{o}-Carretero et al.}
%   \date{Received September 15, 1996; accepted March 16, 1997}

% \abstract{}{}{}{}{} 
% 5 {} token are mandatory
 
  \abstract
   {Stellar tidal streams are the result of tidal interactions between a central galaxy and lower mass systems like satellite galaxies or globular clusters. For the Local Group, many diffuse substructures have been identified and their link to the galaxy evolution has been traced. However it cannot be assumed that the Milky Way or M31 are representative of their galaxy class, and a larger sample of analogue galaxies beyond the Local Group is required to be able to generalise the underlying theory.}
   {We want to characterise photometrically the stellar streams around Milky Way analogues in the local Universe with the goal to deepen our understanding of the interaction between host and satellite galaxies, and ultimately of the galaxy formation and evolution processes. }
   {In the present work we identified and analysed stellar tidal streams around Milky Way analogue galaxies from the SAGA sample, using deep images of the DESI Legacy Imaging Surveys (for this sample, we obtain a range of $r$-band surface brightness limit between 27.8 and $29\, \mathrm{mag\, arcsec}^{-2}$). We measure the surface brightness and colours of the detected streams using GNU Astronomy Utilities software.}
   {We identified 16 new stellar tidal streams around Milky Way analogue galaxies at distances between 25 and 40 Mpc. Applying statistical analysis to our findings for the SAGA II galaxy sample, we obtained a frequency of 12.2\% $\pm$ 2.4\% for stellar streams. We measured surface brightness and colours of the detected streams, and the comparison to the dwarf satellite galaxies population around galaxies belonging to the same SAGA sample shows that the mean colour of the streams is 0.20~mag redder than that of the SAGA satellites; also, the streams are, in average, $0.057 \pm 0.021$~mag redder that their progenitor, for those cases when a likely progenitor could be identified.}


   \maketitle
%
%-------------------------------------------------------------------

\section{Introduction}
\label{sec:introduction}

 Over the last two decades, studies focused on the formation and evolution of our Galaxy have been significantly advanced by the first generation of wide-field, digital imaging surveys and
 the Gaia astrometric mission. The extensive photometric databases that resulted have provided, for the first time, spectacular panoramic views of the Milky Way tidal streams \citep{belokurov_2006,ibata_2007,ibata2019a,mcconnachie_2009,shipp_2018} and revealed the existence of  large stellar sub-structures in the halo, which have been interpreted as observational evidence of our home Galaxy's hierarchical formation. Furthermore, the PAndAS Survey \citep{mcconnachie_2009} has revealed a panoramic view of the Andromeda halo with a multitude of tidal streams, arcs, shells and other irregular structures that are possibly related to ancient merger events. These observations confirm the $\Lambda$CDM prediction that tidally disrupted dwarf galaxies are important contributors to the formation of Galactic stellar halos. The next generation of Galactic and extragalactic surveys (e.g.\ LSST) will dissect the stellar halo structure of these Local Group spirals with unprecedented detail, promising further improvements in our understanding of the early formation and merger history of the Milky Way.

While some of the known Milky Way and M31 stellar streams can be well characterized in a wide parameter space and also using observations of their individual stars, results for individual systems are not easy to compare with numerical simulations due to the natural stochasticity of galaxy assembly histories in the $\Lambda$CDM model. Although statistical distributions, for example of halo assembly times or satellite luminosities, are well-defined for galaxies selected in a narrow range of stellar mass and/or halo mass, individual systems may show large deviations from the mean. To overcome this limitation, a search for 
%analogs to these galactic fossils 
streams and other merger debris in a larger sample of Milky Way-like galaxies is required.
%to understand if the recent merger histories of the Local Group spirals are ‘typical’. 
This is a daunting task. Because of their extremely faint surface brightness, the observed frequency of stellar streams is very low even in ultra-deep imaging surveys; see \citet{hood_2018} for a modern review. 

In this paper, we will focus only on {\it stellar tidal streams}, arising from the tidal disruption of dwarf galaxies by more massive systems.  
We exploit the deep, wide-field imaging from the DESI Legacy Surveys  \citep{dey2019} to systematically explore the frequency and photometric properties of streams in the stellar halos of 181 Milky Way analogue targets previously selected for the {\it Satellites Around Galactic Analogs} (SAGA) survey \citep{geha2017,mao2021}. 
 
\begin{figure}
\centering
  \includegraphics[width=0.85 \columnwidth]{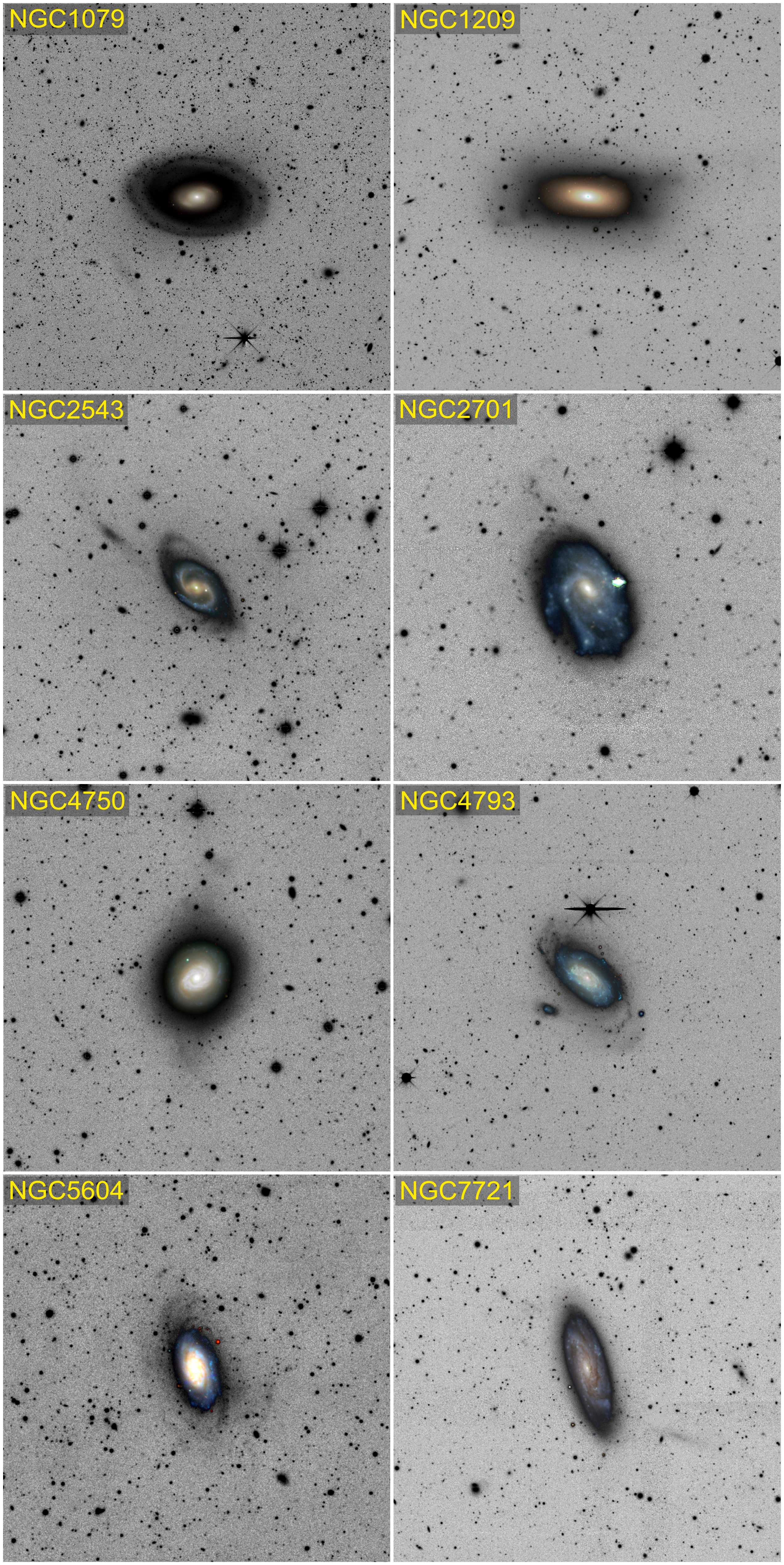}
  \caption{Sample of images showing stellar streams around galaxies listed in Table \ref{tab:photometry}. For illustrative purposes, shallower colour images (also from the {\it DESI Legacy Imaging Surveys}) have been superimposed on saturated central region of each host galaxy.}
  \label{fig-sample}
\end{figure}

\begin{table*}
\centering
{\small
\caption{ Photometry of stellar streams around MW analogue galaxies. Column 1 gives the name of the host galaxy and column 2 its distance; column 3 shows the surface brightness limit in the $r$ band calculated in this work; columns 3 and 4 show the {\it Detection Significance Index}, as defined in \citet{martinez-delgado2021}. Columns 5 to 7 show the surface brightness in the $g$ passband, in the $r$ passband, and the $(g - r)_\mathrm{0}$ colour of the streams, averaged over all the apertures placed on the stream; column 8 indicates whether the stream has been reported for the first time in {\it this work}, indicated by ($\ast$), or in one of the following previous works: (1) \citet{martinez-delgado2021}; (2) \citet{morales2018}; (3) \citet{ludwig2014}; (4) \citet{knierman2013}}.
\label{tab:photometry}

\begin{tabular}{lcccccccc}

Host & D & $\mu_\mathrm{r, limit}$ & \multicolumn{2}{c}{$\mathrm{DSI}_\mathrm{stream}$} & $\langle \mu_{g}\rangle_\textrm{stream}$ & $\langle\mu_{r}\rangle_\mathrm{stream}$ & $\langle (g - r)_\mathrm{0} \rangle_\mathrm{stream}$ & Reference   \\

 & & & maximum & average & & & &  \\
 
  & Mpc & [mag arcsec$^{-2}$]  & $\sigma$ & $\sigma$ & [mag arcsec$^{-2}$] & [mag arcsec$^{-2}$] & [mag] &   \\

\hline\hline

NGC0636	  & 29.2 &    28.88 &	45.58 &	31.86 & 26.66	$\pm$ 0.03 & 25.86	$\pm$ 0.02 &  0.75 	$\pm$ 0.04 & ($\ast$)      \\
NGC1079	  & 31.4 &    28.78 &	15.24 &	11.31 &	27.51	$\pm$ 0.05 & 27.00 	$\pm$ 0.05 &  0.48 	$\pm$ 0.07 & ($\ast$)      \\
%NGC1084	  &    28.92 &	32.42 &	19.28 &	27.14	$\pm$ 0.03 & 26.61	$\pm$ 0.03 &  0.52	$\pm$ 0.04 & (1)       \\
%NGC1097	  &    28.86 &	unrel.&	unrel.& 27.07	$\pm$ 0.03 & 26.38	$\pm$ 0.03 &  0.70	$\pm$ 0.04 & ($\ast$)      \\
NGC1209	  & 38.3 &    28.91 &    8.85 &	4.71  & 28.71	$\pm$ 0.05 & 27.98 	$\pm$ 0.03 &  0.68 	$\pm$ 0.07 & ($\ast$)	   \\
NGC1309	  & 34.3 &    28.76 &	24.42 &	23.02 &	25.66	$\pm$ 0.02 & 26.26 	$\pm$ 0.02 &  0.56 	$\pm$ 0.02 & (1)	   \\
NGC2460   & 34.8 &    28.81 &	10.39 &	8.06  &	27.50	$\pm$ 0.05 & 26.57 	$\pm$ 0.04 &  0.85 	$\pm$ 0.02 & (3)	   \\
NGC2543	  & 37.6 &    28.55 &	10.18 &	9.00  &	26.66	$\pm$ 0.06 & 25.86	$\pm$ 0.06 &  0.72	$\pm$ 0.08 & ($\ast$) 	   \\
NGC2648   & 32.7 &    28.19 &	22.70 & 16.62 &	26.49	$\pm$ 0.03 & 25.96	$\pm$ 0.04 &  0.49	$\pm$ 0.05 & ($\ast$)	   \\
NGC2701	  & 36.5 &    28.58 &	6.63  &	5.55  &	26.85	$\pm$ 0.07 & 26.47	$\pm$ 0.08 &  0.37	$\pm$ 0.10 & ($\ast$)	   \\
NGC2782	  & 39.9 &    28.51 &	28.69 &	20.55 &	26.14	$\pm$ 0.01 & 25.63	$\pm$ 0.02 &  0.48	$\pm$ 0.02 & (4)	   \\
%NGC3583   &    28.36 &	9.94  &	7.13  &	27.00	$\pm$ 0.02 & 26.55	$\pm$ 0.04 &  0.45	$\pm$ 0.04 & ($\ast$)	   \\ major merger?
NGC3614	  & 36.1 &    28.57 &	9.79  &	6.64  &	27.78	$\pm$ 0.06 & 27.07	$\pm$ 0.05 &  0.68	$\pm$ 0.08 & ($\ast$)	   \\      
NGC3689	  & 39.8 &    28.00 &	10.75 &	6.45  &	27.55	$\pm$ 0.05 & 26.82	$\pm$ 0.05 &  0.56  $\pm$ 0.07 &    (1)	   \\
%NGC4203	  &    28.47 &	29.89 &	22.71 &	25.64	$\pm$ 0.02 & 25.13	$\pm$ 0.02 &  0.51	$\pm$ 0.02 &    (3), (4)   \\
NGC4378	  & 37.2 &    28.21 &	24.06 & 22.17 &	27.24	$\pm$ 0.03 & 26.53	$\pm$ 0.03 &  0.68	$\pm$ 0.04 & ($\ast$)	   \\
%NGC4414	  &    28.00 &	10.46 &	8.70  &	28.04	$\pm$ 0.15 & 26.75	$\pm$ 0.09 &  unreliable &          (5)    \\
NGC4750	  & 27.7 &    28.57 &	54.58 &	35.07 &	26.81	$\pm$ 0.02 & 26.30	$\pm$ 0.03 &  0.48	$\pm$ 0.03 & ($\ast$)	   \\    
NGC4793	  & 36.3 &    28.11 &	20.02 &	18.04 &	26.16	$\pm$ 0.04 & 25.60	$\pm$ 0.06 &  0.55	$\pm$ 0.07 & ($\ast$)	   \\
NGC4799	  & 40.1 &    27.93 &	8.49  &	6.98  &	26.65	$\pm$ 0.04 & 26.20	$\pm$ 0.07 &  0.41	$\pm$ 0.08 & ($\ast$)	   \\
%NGC4866   &    28.17 &	11.94 &	9.32  &	25.88	$\pm$ 0.05 & 25.70	$\pm$ 0.08 &  0.18	$\pm$ 0.09 &    (4)	   \\ stream?
NGC5297	  & 35.5 &    28.55 &	28.00 &	18.58 &	26.35	$\pm$ 0.04 & 25.70	$\pm$ 0.04 &  0.63	$\pm$ 0.05 & ($\ast$)	   \\
NGC5493	  & 40.05 &    28.30 &	32.96 &	28.06 &	26.38	$\pm$ 0.02 & 25.69	$\pm$ 0.02 &  0.63	$\pm$ 0.003 & ($\ast$)	   \\
NGC5604	  & 39.0 &    28.18 &	12.29 &	9.93  &	26.35	$\pm$ 0.05 & 25.81	$\pm$ 0.05 &  0.46	$\pm$ 0.07 & ($\ast$)	   \\
NGC5631	  & 31.7 &    28.54 &	12.88 &	10.01 &	27.60	$\pm$ 0.04 & 26.98	$\pm$ 0.04 &  0.59	$\pm$ 0.06 & ($\ast$)	   \\
NGC5750	  & 25.3 &    28.23 &	29.41 &	27.37 &	27.38	$\pm$ 0.05 & 26.69	$\pm$ 0.04 &  0.63	$\pm$ 0.06 &    (2)    \\
NGC5812	  & 27.2 &    28.38 &	55.09 &	30.73 &	26.54	$\pm$ 0.04 & 25.67	$\pm$ 0.02 &  0.77	$\pm$ 0.04 & ($\ast$)	   \\
NGC7721	  & 31.8 &    27.87 &	19.44 &	13.24 &	25.79	$\pm$ 0.03 & 25.23	$\pm$ 0.04 &  0.53	$\pm$ 0.04 & (3)	   \\

\hline

\end{tabular}
}
\end{table*}

\section{Methodology}
\label{sec:methodology}

\subsection{Image Sample}
\label{sec:imagesample}

The second phase of the SAGA survey \citep{mao2021} defines a parent sample of Milky Way-like host galaxies with absolute $K$-band magnitude in the range $-23 < M_{K} < -24.6$ mag, approximately equivalent to the stellar mass range $10^{10} < M_{\star} < 10^{11}\,\mathrm{M}_{\odot}$. The sample excludes close pairs of hosts, defined by a host-satellite K-band magnitude difference of $\Delta K < 1.6$ mag. The SAGA survey only carried out spectroscopic follow-up for hosts in this parent sample with distances $25 < d < 40.75$~Mpc. 
%Here we base our study on the full SAGA II parent sample, including galaxies within $d < 25$~Mpc, which therefore comprises 226 Milky Way analogs with distances $d < 40.75$ Mpc. 
Further details of the SAGA II parent sample can be found in \citet{mao2021}.

We inspected the images of the resulting sample of 181 galaxies using the Legacy Survey Sky Viewer\footnote{\url{https://www.legacysurvey.org/viewer}} and selected for further analysis a subset of targets in which stellar tidal streams could be identified by eye. From this visual inspection, a total of 22 galaxies with detected streams were selected. Image cutouts of these selected targets were then computed from the raw data from the {\it DESI Legacy Imaging Surveys} \citep[][; LS]{dey2019} using a modified version of the LS reduction pipeline {\it Legacypipe}. This alters the way the image backgrounds (``sky models'') are computed; {\it Legacypipe} by default uses a flexible spline sky model which can over-subtract the outskirts of large galaxies.  Instead, we assume a flat background level for each overlapping CCD.  We first minimize the relative background levels between the overlapping CCDs in each band, and then, after detecting and masking sources as well as Gaia stars, we subtract the sigma-clipped median in the outer half of the image cutout (see \citet[][]{martinez-delgado2021} for details). The resulting wide-field images reach surface brightness limits as faint as 29 mag arcsec$^{-2}$ in the $r$ band (see Section \ref{sec:dataanalysis}), ensuring a sufficient image depth to be able to measure very faint tidal structures. The images  analysed in this work are listed in Table \ref{tab:photometry} and examples of them are shown in Figure \ref{fig-sample}.

\begin{figure}[h!]
\centering
\includegraphics[width=0.8\columnwidth]{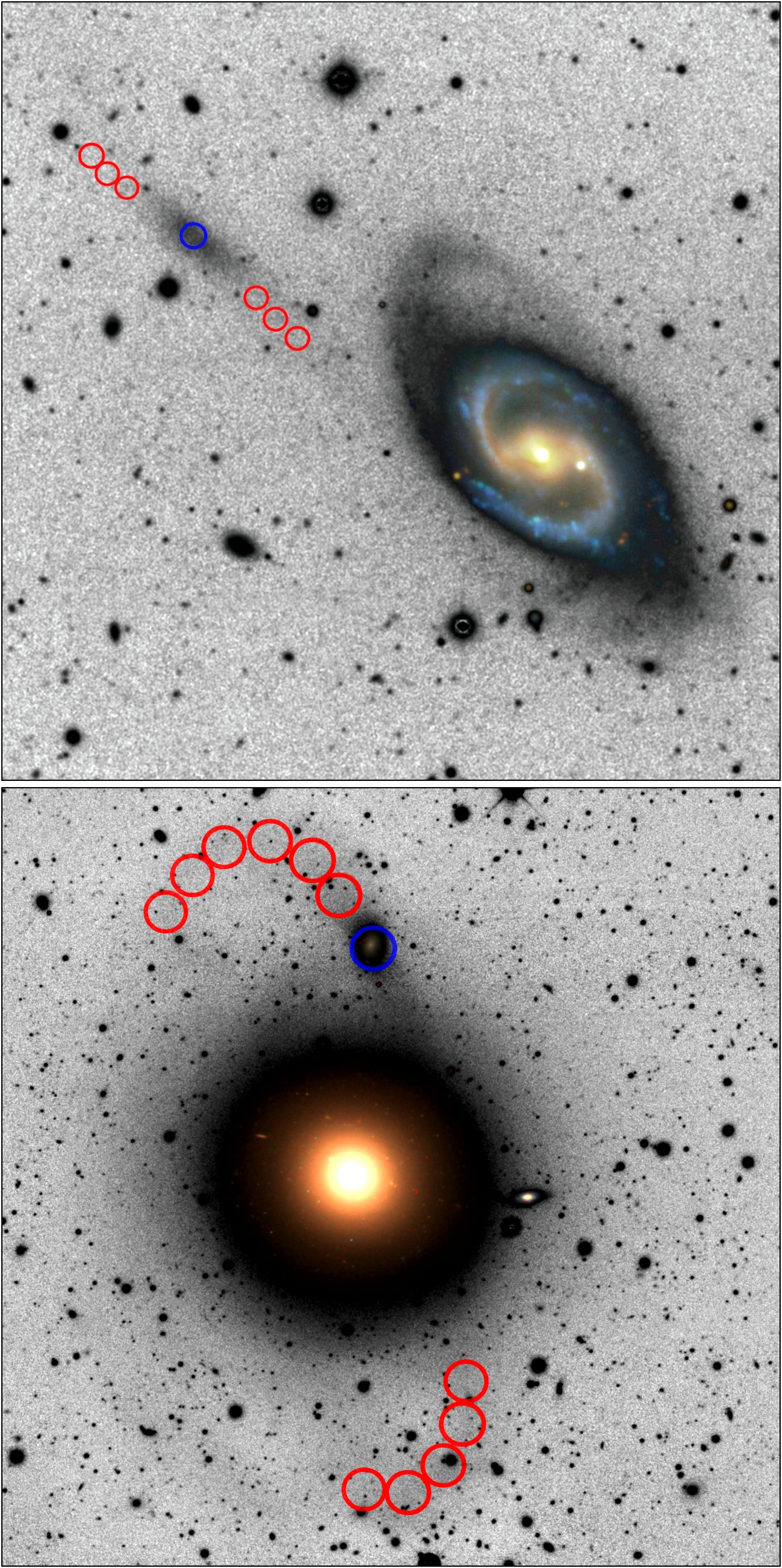}
  \caption{Examples of our photometry measurement method, showing the apertures placed on the stellar streams around NGC5812 and NGC2543 along with the suspected progenitors, in order to measure their surface brightness and colours.} 
  \label{fig-photometry}
\end{figure}

\begin{figure}
     \includegraphics[width=\columnwidth]{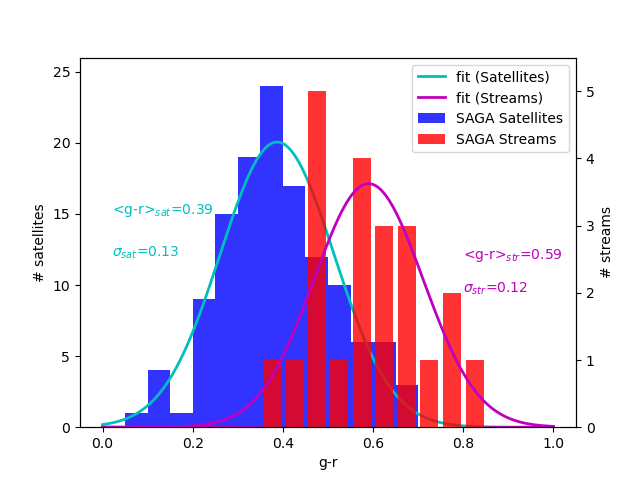}
    \caption{Histogram showing the distribution of the average $(g - r)_\mathrm{0}$ colour of stellar streams around 22 galaxies from our sample, (those listed in Table~\ref{tab:photometry}) together with the same colour of the 127 satellite galaxies from the 36 SAGA systems sample.}
    \label{fig-histograms}
\end{figure}

\begin{table}
\centering
{\small
\caption{Comparison between the average $(g - r)_\mathrm{0}$ colour of each streams and the corresponding colour of its visually identified progenitor.}
\label{tab:colors}

\begin{tabular}{lccc}

Host & $\langle (g - r)_\mathrm{0} \rangle_\mathrm{stream}$ & $\langle (g - r)_\mathrm{0} \rangle_\mathrm{progenitor}$ & $\Delta$ \\
   & [mag] & [mag] & [mag] \\
\hline\hline

NGC2543	& 0.72	$\pm$0.08 &	0.51	$\pm$ 0.02 & 0.21 $\pm$ 0.08 \\
NGC2648	& 0.49	$\pm$0.05 &	0.56	$\pm$0.003 & -0.07 $\pm$ 0.05\\
NGC3614	& 0.68	$\pm$0.08 &	0.65	$\pm$0.08  & 0.03 $\pm$ 0.11\\
NGC3689	& 0.56	$\pm$0.07 &	0.59	$\pm$0.02  & -0.03 $\pm$ 0.07 \\
NGC4793	& 0.55	$\pm$0.07 &	0.39	$\pm$0.01  & 0.16 $\pm$ 0.07\\
NGC5297	& 0.63	$\pm$0.05 &	0.64	$\pm$0.004 & -0.01 $\pm$ 0.05\\
NGC5750	& 0.63	$\pm$0.06 &	0.57	$\pm$0.02  & 0.06 $\pm$ 0.06\\
NGC5812	& 0.77	$\pm$0.04 &	0.63	$\pm$0.005 & 0.14 $\pm$ 0.04\\

\hline

\end{tabular}
}
\end{table}

\subsection{Data Analysis}
\label{sec:dataanalysis}

We carried out the photometric analysis with {\it GNU Astronomy Utilities} (Gnuastro)\footnote{\url{http://www.gnu.org/software/gnuastro}}. 
We made all the measurements by applying Gnuastro's {\sc MakeCatalog} subroutine on the sky-subtracted images generated by Gnuastro's {\sc NoiseChisel} \citep{Akhlaghi15,Akhlaghi19}. The program also provides us with the errors in the photometry, calculated as inversely proportional to the SNR \footnote{\url{https://www.gnu.org/software/gnuastro/manual/html_node/Magnitude-measurement-error-of-each-detection.html}}.

Our photometric analysis includes  measurements of surface brightness in the LS $r$ and $g$ passbands for each stream, and for their candidate progenitor satellite, where identified. Taking advantage of the depth and photometric quality of the LS survey images, we have also measured the $(g - r)_\mathrm{0}$ colour of the streams.  
We measure the surface brightness limit of the images for the $g$ and $r$ passbands following the approach of \cite{Roman2020}, i.e. we report the value corresponding to $+3\sigma$ of the sky background in an area of 100 arcsec$^2$. Table~\ref{tab:photometry} reports the surface brightness limit for the $r$ band, which is representative of the depth of the corresponding images in other bands.

We measured surface brightness and colours on circular apertures, placed manually following closely the detection map of the stream generated by {\sc NoiseChisel}, once all foreground and background sources were masked. A succession of circular apertures allows to measure colour gradients and can easily adapt to the stream contour, though in a few cases where the stream shape so allowed, larger polygonal apertures were used to reduce the measurement error. Regions where the stream surface brightness was judged to be significantly blended with light from the host galaxy were avoided. As an illustration of the method, Figure \ref{fig-photometry} shows an example of a stream on which apertures have been placed manually in order to perform the measurement. We obtain a representative surface brightness and colour for each stream by taking the mean of the corresponding individual aperture measurements.

\section{Results }
\label{sec:results}
 
Table \ref{tab:photometry} shows the results of our photometric analysis. We identified tidal streams around 22 galaxies from the sample of 181 MW analogues. This suggests that 12.2\% $\pm$ 2.4\% of the SAGA II galaxies have a stellar stream in the halo, for a $r$-band surface brightness limit range of our images between 27.8 and $29\, \mathrm{mag\, arcsec}^{-2}$ (see Table \ref{tab:photometry}). This implies that, with 95\% confidence, the percentage of typical SAGA sample halos that have readily observable stellar streams is between 7.4\% and 16.9\%. This result is similar to that reported by \cite{morales2018} for their systematic assessment of the frequency of tidal streams around a different sample of Milky Way-like galaxies in the local Universe. \citeauthor{morales2018} used co-added SDSS DR9  $g$, $r$ and $i$ band images processed using an image-enhancing technique similar to that of \cite{miskolczi2011}, with a typical surface brightness limited of $28.1\ \pm 0.3\ \mathrm{mag~arcsec^{-2}}$. They reported a total of 28 tidal streams from a sample of 297 galaxies, providing a conservative estimate that only $\sim 10\%$ of galaxies show evidence of diffuse features that may be linked to satellite accretion events.

The measured ranges of stream surface brightness are $25.66 < \mu_{g} < 28.71$ and $25.23 < \mu_{r} < 27.98$ $\mathrm{mag\, arcsec}^{-2}$. 
The {\it Detection Significance Index} (DSI), as defined in \citet{martinez-delgado2021}, is calculated by comparing the measurements for a given aperture with the median and standard deviation of $N$ random measurements in pixels with no source detection \footnote{\url{https://www.gnu.org/software/gnuastro/manual/html_node/Upper-limit-magnitude-of-each-detection.html}}. 
The {\it Reference} column of Table \ref{tab:photometry} indicates whether each stream has been reported in the literature or is reported for the first time in this work.

Figure~\ref{fig-histograms} compares the $(g - r)_\mathrm{0}$ colour distribution of the stellar streams identified in Table \ref{tab:photometry}, shown in red, to that of the 127 spectroscopically confirmed satellite galaxies from the 36 SAGA systems presented in \citet{mao2021}, shown in blue. 
Hypothesis contrast of normality shows that the null hypothesis that these colour distributions come from a Gaussian distribution cannot be rejected with a 99\% confidence level. We therefore fit Gaussian functions to each distribution, finding means and standard deviations of $0.59 \pm 0.12\,\mathrm{mag}$ for the streams and $0.39 \pm 0.13 \,\mathrm{mag}$ for the SAGA satellites.  
The mean colour of the streams is therefore 0.20 mag redder than that of the SAGA satellites. An equality of means hypothesis test shows that the null hypothesis can be rejected with a statistical confidence level larger than 99.999\% (p-value $< 10^{-10}$) and the alternative hypothesis that mean colour of the streams is redder than mean colour of satellites can be accepted.
The $(g - r)_\mathrm{0}$ colours we find are similar to those obtained for the streams described in the proof-of-concept study of \citet{martinez-delgado2021}, who reported a mean and standard deviation of $0.66\pm 0.12$~mag.

In approximately $36\%$ of the streams in our sample, we have identified a highly likely progenitor by visual inspection. This allows us to explore similarities and differences in the stellar populations of satellites and their streams, including the presence of population gradients along the streams. 
As shown in Fig.~\ref{fig-photometry} for the cases of NGC 2543 and NGC 5812, we placed apertures on the the likely progenitors as well as along the tidal features. Table~\ref{tab:colors} compares the $(g - r)_\mathrm{0}$ colour of the stream (averaged over the apertures as described in Section \ref{sec:dataanalysis}) with that measured in an aperture placed on the suspected progenitor. We see a significant difference in colour for the streams around NGC2543, NGC4793 and NGC5812, with the stream redder than its likely progenitor by 0.21, 0.16 and 0.14~mag, respectively. For the rest of streams where a progenitor is suspected, the colour difference is within the uncertainties of our colour measurement, and therefore not significant. To test whether the differences observed in our sample are statistically significant or not, we have performed a hypothesis test of the difference between the stream and the progenitor colours, and we have obtained that streams are, on average, $0.057 \pm 0.021$~mag redder that their progenitor, with a confidence level $>99.99\%$.

\section{Conclusions}
\label{sec:discussion}

The main conclusions of this letter are as follows:
\begin{itemize}
\item We have developed a new methodology, based on Gnuastro, for measuring the surface brightness and colours of streams.
\item We have applied this methodology to enhanced DESI Legacy Imaging Survey $grz$ data for a subset of the SAGA sample (a stellar mass-selected sample of Milky Way analogues at distances up to 40 Mpc). 
\item We have detected 16 previously unreported streams in this sample (see table 1, {\it Reference} column). The streams we have analyzed have $r$-band surface brightnesses in the range $25.23 < \mu_{r} < 27.98\,\mathrm{mag\,arcsec^{-2}}$.
\item We have carried out a statistical comparison of $(g - r)_\mathrm{0}$ colours for the detectable stream and satellite populations in our sample, finding that the detectable stream population is significantly redder on average.
\item In those systems where a progenitor of the stream could be identified by visual inspection, we find the stream is on average slightly redder than the progenitor.
\end{itemize}

We suggest that the differences we find between the stream and satellite colour distributions may be explained by a combination of selection bias and physical effects. We provide here a brief summary of possible explanations, and defer a detailed discussion to future work.

The SAGA survey selects a sample of candidate satellites based on catalogue photometry and follows up a subset of these with multi-object fibre spectrographs to obtain redshifts. Extremely compact (M32-like) candidates were not followed up \citep{geha2017}; although such objects tend to be red, relatively few are known. More significantly, redshifts are more difficult to obtain for candidates with low mean surface brightness, which also tend to be redder. \citet{mao2021} argue that this redshift incompleteness is a weak effect that does not significantly bias the distribution of star formation rates (hence colours) in the spectroscopic sample. However, the completeness of the initial target catalogue may also be important. \citet{font22} explore this issue in detail through comparison to the ARTEMIS suite of cosmological simulations. They suggest that the photometric SAGA candidate sample may have a significant bias against low surface brightness satellites, and that this bias has a much stronger effect on the resulting colour distribution. Comparing to a separate survey of satellites in the Local Volume (Exploration of Local VolumE Satellites, {\sc ELVES}, see \citet{carlsten2021}), they find evidence that fainter galaxies in SAGA are biased towards bluer colours.

However, even with the small sample of stream colours presently available, we find at least two reasons to consider physical explanations for the colour differences in addition to selection effects. First, \citet{font22} find the potential selection bias in SAGA mostly affects the fainter satellite magnitudes ($M_{\mathrm{V}}>-12$), and that the colours of brighter (systematically bluer) satellites are not strongly biased. Although we cannot yet quantify the total luminosity of the streams in our sample, it is likely that readily detectable streams have some bias towards the brighter end of the luminosity function of disrupted progenitors (albeit with large uncertainty due to the wide variety of stream morphology and viewing angle). If we were to compare the streams only to the brighter SAGA satellites, rather than the full sample, the discrepancy in colour would be reinforced. Put another way, we detect no streams as blue as the bluest SAGA satellites.

Secondly, the difference in colour seen in the small number of stream-progenitor pairs in our sample suggests colour gradients may contribute alongside selection-driven differences between the stream and satellite samples (and other population-level effects, such as different average ages). Such gradients may be established either before disruption or during the disruption process. A wide variety of physical processes could create gradients through their effects on the relative timescales of gas removal (due to ejection and ram pressure stripping), star formation in residual cold gas, and tidal stripping. At the most basic level, complete tidal disruption will prevent further star formation, leading to the systematic reddening of dynamically older streams. Cosmological simulations are necessary to make quantitative predictions for colour distributions, accounting for the range of satellite star formation histories, gas fractions and orbits, and  variations in the satellite accretion rate and disruption efficiency over the range of dark matter halo masses that may correspond to the SAGA sample.

To make further progress, we are currently constructing a larger sample of galaxies within the Stellar Streams Legacy Survey \citep{martinez-delgado2021}. This sample will comprise more than 800 Milky Way-like galaxies.
%, and with the results of their analysis,
By analysing this sample 
using the techniques presented in this paper,
we will be able to more robustly test our conclusions and carry out meaningful comparisons to physical models of satellite star formation, accretion and disruption. 

\begin{acknowledgements}
        We want to thank to Yao-Yuan Mao, Marla Geha and Risa Wechsler for providing the original SAGA sample for this paper and useful comments.  We also thank Dustin Lang and John Moustakas for running the modified {\it Legacypipe} code to produce the images used here.
        DMD acknowledges financial support from the Talentia Senior Program (through the incentive ASE-136) from Secretar\'\i a General de  Universidades, Investigaci\'{o}n y Tecnolog\'\i a, de la Junta de Andaluc\'\i a. DMD acknowledge funding from the State Agency for Research of the Spanish MCIU through the ``Center of Excellence Severo Ochoa" award to the Instituto de Astrof{\'i}sica de Andaluc{\'i}a (SEV-2017-0709) and project (PDI2020-114581GB-C21/ AEI / 10.13039/501100011033). 
        MAGF acknowledges financial support from the Spanish Ministry of Science and Innovation through the project PID2020-114581GB-C22. SRF acknowledge financial support from the Spanish Ministry of Economy and Competitiveness (MINECO) under grant number AYA2016-75808-R, AYA2017-90589-REDT and S2018/NMT-429, and from the CAM-UCM under grant number PR65/19-22462. SRF acknowledges support from a Spanish postdoctoral fellowship, under grant number 2017-T2/TIC-5592. 
        APC is supported by the Taiwan Ministry of Education Yushan Fellowship and Taiwan National Science and Technology Council grant 109-2112-M-007-011-MY3. 
        The photometry analysis in this work was partly done using GNU Astronomy Utilities (Gnuastro, ascl.net/1801.009) version $0.17$. Work on Gnuastro has been funded by the Japanese MEXT scholarship and its Grant-in-Aid for Scientific Research (21244012, 24253003), the European Research Council (ERC) advanced grant 339659-MUSICOS, and from the Spanish Ministry of Economy and Competitiveness (MINECO) under grant number AYA2016-76219-P.
        The Leiden Observatory has provided facilities and computer infrastructure for carrying out part of this work.
        M.A acknowledges the financial support from the Spanish Ministry of Science and Innovation and the European Union - NextGenerationEU through the Recovery and Resilience Facility project ICTS-MRR-2021-03-CEFCA.
        
\end{acknowledgements}


\begin{thebibliography}{}

    \bibitem[\protect\citeauthoryear{Akhlaghi and Ichikawa}{2015}]{Akhlaghi15} Akhlaghi M., Ichikawa T., 2015, ApJS, 220, 1.
    \bibitem[\protect\citeauthoryear{Akhlaghi}{2019}]{Akhlaghi19} Akhlaghi M., 2019, ASPC, 521, 299A.
     \bibitem[Belokurov et al.(2006)]{belokurov_2006} Belokurov, V., Zucker, D.~B., Evans, N.~W., et al.\ 2006, \apjl, 642, L137
     \bibitem[Carlsten et al.(2021)]{carlsten2021} Carlsten, S.~G., Greene, J.~E., Greco, J.~P., et al.\ 2021, \apj, 922, 267. doi:10.3847/1538-4357/ac2581
    \bibitem[Dey et al.(2019)]{dey2019} Dey, A., Schlegel, D.~J., Lang, D. et al.\ 2019, \aj, 157, 168. doi:10.3847/1538-3881/ab089d
    \bibitem[Font et al.(2022)]{font22} Font, A.~S., McCarthy, I.~G., Belokurov, V., et al.\ 2022, \mnras, 511, 1544. doi:10.1093/mnras/stac183
    \bibitem[Geha et al.(2017)]{geha2017} Geha, M., Wechsler, R.~H., Mao, Y.-Y., et al.\ 2017, \apj, 847, 4. doi:10.3847/1538-4357/aa8626
    \bibitem[Hood et al.(2018)]{hood_2018} Hood, C.~E., Kannappan, S.~J., Stark, D.~V., et al.\ 2018, \apj, 857, 144. doi:10.3847/1538-4357/aab719
    \bibitem[Ibata et al.(2007)]{ibata_2007} Ibata, R., Martin, N.~F., Irwin, M., et al.\ 2007, \apj, 671, 1591
    \bibitem [Ibata(2019)]{ibata2019a} Ibata, R., Malhan, K., \& Martin, N. 2019a, arXiv e-prints, https://arxiv.org/abs/1901.07566
    \bibitem[Knierman et al.(2013)]{knierman2013} Knierman, K.~A., Scowen, P., Veach, T., et al.\ 2013, \apj, 774, 125. doi:10.1088/0004-637X/774/2/125
    \bibitem[Ludwig (2014)]{ludwig2014} Ludwig, J. 2014, Ph. D. thesis, Universität Heidelberg
    \bibitem[Mao et al.(2021)]{mao2021} Mao, Y.-Y., Geha, M., Wechsler, R.~H., et al.\ 2021, \apj, 907, 85. doi:10.3847/1538-4357/abce58
    \bibitem[Martinez-Delgado et al.(2021)]{martinez-delgado2021} Martinez-Delgado, D., Cooper, A.~P., Roman, J., et al.\ 2021, arXiv:2104.06071
     \bibitem[McConnachie et al.(2009)]{mcconnachie_2009} McConnachie, A.~W., Irwin, M.~J., Ibata, R.~A., et al.\ 2009, \nat, 461, 66
     \bibitem[Miskolczi et al.(2011)]{miskolczi2011} Miskolczi, A., Bomans, D.~J., \& Dettmar, R.-J.\ 2011, \aap, 536, A66. doi:10.1051/0004-6361/201116716
     \bibitem[Morales et al.(2018)]{morales2018} Morales, G., Mart{\'\i}nez-Delgado, D., Grebel, E.~K., et al.\ 2018, \aap, 614, A143. doi:10.1051/0004-6361/201732271
    \bibitem[Rom{\'a}n et al.(2020)]{Roman2020} Rom{\'a}n, J., Trujillo, I., \& Montes, M.\ 2020, \aap, 644, A42
    \bibitem[Shipp et al.(2018)]{shipp_2018} Shipp, N., Drlica-Wagner, A., Balbinot, E., et al.\ 2018, \apj, 862, 114


\end{thebibliography}
\end{document}